\RequirePackage{fix-cm}

\documentclass[twocolumn]{svjour3}

\newcommand{\comm}[1]{}

\smartqed

\usepackage{graphicx,amssymb}
\usepackage{epsf,psfig}
\usepackage{txfonts}
\usepackage{natbib}

\usepackage[dvipdfm=true]{hyperref}
\hypersetup{pdftitle = The title of my PDF, pdfauthor = My name, pdfsubject= The subject, pdfkeywords = keyword1 keyword2 keyword3}
\hypersetup{colorlinks = true, linkcolor = blue, anchorcolor = red, citecolor = blue, filecolor = red, pagecolor = red, urlcolor = blue}

\journalname{Astrophysics and Space Science}

\begin{document}

\title{Orbital dynamics in the planar Saturn-Titan system}

\author{Euaggelos E. Zotos}

\institute{Department of Physics, School of Science, \\
Aristotle University of Thessaloniki, \\
GR-541 24, Thessaloniki, Greece \\
Corresponding author's email: {evzotos@physics.auth.gr}}

\date{Received: 10 April 2015 / Accepted: 25 May 2015 / Published online: 3 June 2015}

\titlerunning{Orbital dynamics in the planar Saturn-Titan system}

\authorrunning{Euaggelos E. Zotos}

\maketitle

\begin{abstract}
We use the planar circular restricted three-body problem in order to numerically investigate the orbital dynamics of orbits of a spacecraft, or a comet, or an asteroid in the Saturn-Titan system in a scattering region around the Titan. The orbits can escape through the necks around the Lagrangian points $L_1$ and $L_2$ or collide with the surface of the Titan. We explore all the four possible Hill's regions depending on the value of the Jacobi constant. We conduct a thorough numerical analysis on the phase space mixing by classifying initial conditions of orbits and distinguishing between three types of motion: (i) bounded, (ii) escaping and (iii) collisional. In particular, we locate the different basins and we relate them with the corresponding spatial distributions of the escape and crash times. Our results reveal the high complexity of this planetary system. Furthermore, the numerical analysis shows a strong dependence of the properties of the considered basins with the total orbital energy, with a remarkable presence of fractal basin boundaries along all the regimes. We hope our contribution to be useful in both space mission design and study of planetary systems.


\end{abstract}

\section{Introduction}
\label{intro}

Over the years many research works have been devoted on the issue of escaping or leaking particles from several types dynamical systems. In particular, the problem of escape is a classical problem in Hamiltonian nonlinear systems \citep[e.g.,][]{AVS01,AS03,AVS09,BBS08,BSBS12,Z14} as well as in dynamical astronomy \citep[e.g.,][]{HB83,BGOB88,BTS96,BST98,dML00,Z12}, while escaping orbits in the classical Restricted Three-Body Problem (RTBP) is another typical example \citep[e.g.,][]{N04,N05,DH11,DH12,dAT14,Z15b}.

Nevertheless, the issue of escaping orbits in Hamiltonian systems is by far less explored than the closely related problem of chaotic scattering \citep{BGOB88}. In this situation, a test particle coming from infinity approaches and then scatters off a complex potential. In a recently published paper \citep{Z15b} we investigated the orbital dynamics in the Copenhagen problem where one of the two primary bodied was a oblate spheroid. There the scattering region on the configuration $(x,y)$ space was a rectangular region extending around both primaries and we managed to classified initial conditions of orbits into different categories (i.e., orbits that crash into one of the primaries, escaping orbits and bounded orbits of several types). In the current work on the other hand, the scattering region of initial conditions of orbits is confined only around the Titan and especially between the Lagrangian points $L_1$ and $L_2$.

Escaping and trapped motion of stars in stellar systems is an another issue of great importance. In a previous article \citep{Z12}, we explored the nature of orbits of stars in a galactic-type potential, which can be considered to describe local motion in the meridional $(R,z)$ plane near the central parts of an axially symmetric galaxy. It was observed, that apart from the trapped orbits there are two types of escaping orbits, those which escape fast and those which need to spend vast time intervals inside the equipotential surface before they find the exit and eventually escape. The escape dynamics and the dissolution process of a star cluster embedded in the tidal field of a parent galaxy was investigated in \citet{EJSP08}. Conducting a thorough scanning of the available phase space the authors managed to obtain the basins of escape and the respective escape rates of the orbits, revealing that the higher escape times correspond to initial conditions of orbits near the fractal basin boundaries. The investigation was expanded into three dimensions in \citet{Z15a} where we revealed the escape mechanism of three-dimensional orbits in a tidally limited star cluster. Furthermore, \citet{EP14} explored the escape dynamics in the close vicinity of and within the critical area in a two-dimensional barred galaxy potential, identifying the escape basins both in the phase and the configuration space. The numerical approach of the above-mentioned papers served as the basis of this work.

Many times in the past the RTBP has played an essential role in several areas of celestial mechanics and dynamical astronomy. For instance, the modern applications to space flight missions \citep{KLMR00b,G01a,G01b,G01c,G01d,G04,SYC08} are more numerous than the theoretical classical applications. Moreover, today many issues in space dynamics are of paramount importance and interest. The applications of RTBP include the launching of artificial satellites in the solar system and they also form the basis of several planetary theories.

Titan, Saturn's largest moon, is undoubtedly a treasure trove of potential space discovery. Having an environment which is comparable to that of the primordial Earth and an atmosphere denser than that of any other moon in the solar system makes it a perfect candidate for space missions. In particular,the Cassini-Huygens mission's recent discovery of hydrocarbon lakes on the Titan's surface \citep{KLMR00a} has so inspired the National Aeronautics and Space Administration (NASA) that they plan to deploy a space mission to Titan once more in roughly one decade from now. In the present contribution we shall explore the orbital dynamics of the Saturn-Titan system applying as a guide the numerical approach used in \citet{dAT14} (hereafter Paper I) for the Earth-Moon system.

The structure of the paper is as follows: In Section \ref{dynmod} we present in detail the properties of the dynamical model. All the fundamental concepts and the computational methods we used in order to determine the classification of the orbits are described in Section \ref{cometh}. In the following Section, we conduct a thorough and systematic numerical investigation revealing the overall orbital structure of the Saturn-Titan planetary system. Our paper ends with Section \ref{disc} where the discussion and the main conclusions of this work are given.

\section{The dynamical model}
\label{dynmod}

Let us briefly recall the basic properties of the planar circular restricted three–body problem (PCRTBP) \citep{S67}. The two main bodies, called primaries $P_1$ and $P_2$ move on circular orbits around their common center of gravity. The third body (test particle) moves in the same plane under the gravitational field of the two primaries. It is assumed that the motion of the two primaries is not perturbed by the third body since the third body's mass is much smaller with respect to the masses of the two primaries. The non-dimensional masses of the two primaries are $1-\mu$ and $\mu$, where $\mu = m_2/(m_1 + m_2)$ is the mass ratio and $m_1 > m_2$. For the Saturn-Titan (ST) system, $\mu = 0.0002461294$ \citep{SSR76,BS12}. We choose as a synodic barycenter reference frame a rotating coordinate system where the origin is at the center of mass of $P_1 - P_2$, while the centers $C_1$ and $C_2$ of the two primaries are located at $(-\mu, 0)$ and $(1-\mu,0)$, respectively.

The total time-independent effective potential function in the rotating frame is
\begin{equation}
\Omega(x,y) = \frac{(1 - \mu)}{r_1} + \frac{\mu}{r_2} + \frac{(1 - \mu)A_1}{2r_1^3} + \frac{n^2}{2}\left( x^2  + y^2 \right),
\label{pot}
\end{equation}
where
\[
r_1 = \sqrt{\left(x + \mu\right)^2 + y^2},
\]
\[
r_2 = \sqrt{\left(x + \mu - 1\right)^2 + y^2},
\]
\begin{equation}
n^2 = 1 + \frac{3 A_1}{2},
\label{dist}
\end{equation}
are the distances to the respective primaries and the angular velocity $(n)$.

The larger primary body $P_1$ (Saturn) is an oblate spheroid \citep{BPC99} and $A_1$ is the oblateness coefficient which is defined as
\begin{equation}
A_1 = \frac{(RE)^2 - (RP)^2}{5R^2},
\label{obl}
\end{equation}
where $RE$ and $RP$ are the equatorial and polar radius, respectively of the oblate primary, while $R$ is the distance between the centers of the two primaries. According to Table 1 of \citet{SSR76} the oblateness coefficient for the Saturn-Titan system is $A_1 = 0.0000981153$. The study of oblateness includes the series of works of \citet{MRVK00,KMP05,KDP06,KPP08,KGK12,PPK12,SSR79,SSR86,SRS88,SRS97,S81,S87,S89,S90} by considering the more massive primary as an oblate spheroid with its equatorial plane co-incident with the plane of motion of the primaries.

The scaled equations of motion describing the motion of the third body in the corotating frame read \citep{BS12}
\begin{equation}
\ddot{x} - 2n\dot{y} = \frac{\partial \Omega}{\partial x},  \ \ \ \ddot{y} + 2n\dot{x} = \frac{\partial \Omega}{\partial y}.
\label{eqmot}
\end{equation}
The dynamical system (\ref{eqmot}) admits the well know Jacobi integral
\begin{equation}
J(x,y,\dot{x},\dot{y}) = 2\Omega(x,y) - \left(\dot{x}^2 + \dot{y}^2 \right) = C,
\label{ham}
\end{equation}
where $\dot{x}$ and $\dot{y}$ are the velocities, while $C$ is the Jacobi constant which is conserved and defines a three-dimensional invariant manifold in the total four-dimensional phase space. Therefore, an orbit with a given value of it's energy integral is restricted in its motion to regions in which $C \leq 2 \Omega(x,y)$, while all other regions are forbidden to the third body. The energy value $E$ is related with the Jacobi constant by $C = - 2E$.

\begin{figure}[!tH]
\includegraphics[width=\hsize]{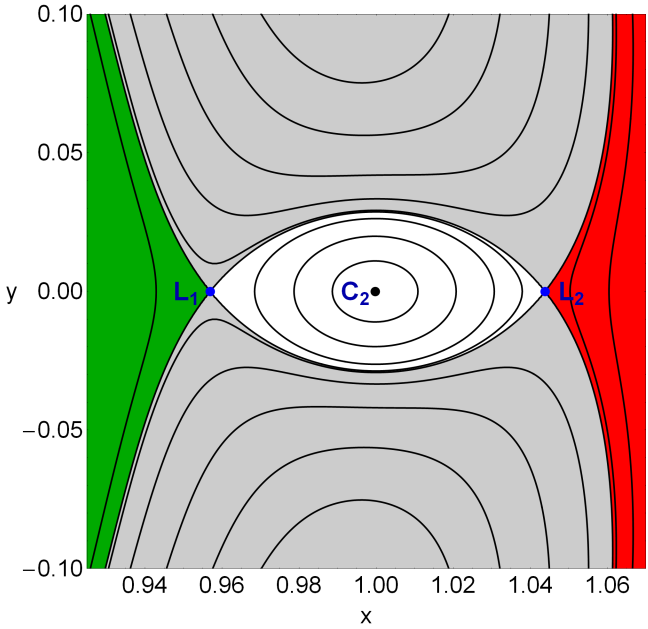}
\caption{The isolines contours of the constant potential and the location of the Lagrangian points around smaller primary (Titan). The interior region around Titan is indicated in white, the region around Saturn is colored in green, the exterior region is shown in red, while the forbidden regions of motion are marked with grey.}
\label{conts}
\end{figure}

The PCRTBP with oblateness has five equilibria known as Lagrangian points at which
\begin{equation}
\frac{\partial \Omega}{\partial x} = \frac{\partial \Omega}{\partial y} = 0.
\label{lps}
\end{equation}
Three of them, $L_1$, $L_2$, and $L_3$, are collinear points located in the $x$-axis, while the other two $L_4$ and $L_5$ are called triangular points and they are located in the vertices of an equilateral triangle. The central stationary point $L_1$ is a local minimum of the potential function $\Omega(x,y)$, while the stationary points $L_2$ and $L_3$ are saddle point. $L_1$ is between the primaries, $L_2$ is at the right side of $P_2$, while $L_3$ is at the left side of $P_1$. The points $L_4$ and $L_5$ on the other hand, are local maxima of the potential function, enclosed by the banana-shaped isolines. The projection of the four-dimensional phase space onto the configuration (or position) space $(x,y)$ is called the Hill's regions. The boundaries of these Hill's regions are called Zero Velocity Curves (ZVCs) because they are the locus in the configuration $(x,y)$ space where the kinetic energy vanishes. The Jacobi constant values at $L_k$ are denoted by $C_k$. For the Saturn-Titan system we have: $C_1 = 3.0164219315$, $C_2 = 3.0160936366$, $C_3 = 3.0004914646$, and $C_4 = C_5 = 2.9999991590$. In Fig. \ref{conts} we see that the presence of forbidden regions in the PCRTBP permits the definition of four subsets of the configuration $(x,y)$ plane when $C_2 < C < C_3$: the interior region around Titan (white), the region around Saturn (green), the exterior region (red), and the forbidden regions (gray).

\section{Computational methods}
\label{cometh}

For exploring the orbital dynamics in the Saturn-Titan system, we need to define samples of initial conditions of orbits whose properties will be identified. For this purpose we consider dense, uniform grids of $1024 \times 1024$ initial conditions $(x_0, y_0)$ regularly distributed on the configuration $(x,y)$ plane inside the area allowed by the Jacobi constant $C$. Following a typical approach, all orbits are launched with initial conditions inside a certain region, called scattering region, which in our case is $x_{L_1} \leq x \leq x_{L_2}$ and $-0.1 \leq y \leq 0.1$. For all orbits $\dot{x_0} = 0$, while the value of $\dot{y_0}$ is always obtained from the Jacobi integral (\ref{ham}) as $\dot{y_0} = \dot{y}(x_0,y_0,\dot{x_0},C) > 0$.

In the PCRTBP system the configuration space extends to infinity thus making the identification of the type of motion of the third body for specific initial conditions a rather demanding task. Depending on the Jacobi constant the region around the Titan can be connected or not to the Saturn realm through the neck around $L_1$ or to the exterior realm through the neck around $L_2$. Therefore initial conditions of orbits around the Titan can be categorized into basins\footnote{The set of initial conditions of orbits which lead to a certain final state (escape, collision or bounded motion) is defined as a basin.}, namely: (i) the bounded basin containing orbits which remain in the Titan realm, (ii) the Saturn realm basin constituted by orbits which escape through $L_1$, (iii) the exterior realm basin corresponding to orbits that escape through $L_2$, (iv) the collisional basin or orbits which collide with the Titan.

Now we need to define appropriate numerical criteria for distinguishing between these four types of motion. In order to consider a more realistic approach, we assume that the Titan is a finite body, taking into account its mean radius approximately by 2576 km (about $6.7 \times 10^{-3}$ dimensionless length units). Therefore, if an orbit reaches the surface of the Titan, its numerical integration ends thus producing an orbit leaking in the configuration space. Furthermore, an escaping orbit to the Saturn realm must satisfy the conditions $x < x_{L_1} - \delta_1$, with $\delta_1 = 0.07$ and the third body inside he circle around the Saturn of radius $r_1 \leq |x_{L_3} - x_{P_1}|$. In the same vein, an escaping orbit to the exterior realm must fulfill the conditions $x > x_{L_2} + \delta_2$, with $\delta_2 = 0.05$, or $|y| > y_{L_5}$, or, if $x < x_{L_1} - \delta_1$, $r_1 > |x_{L_3} - x_{P_1}|$ (i.e., the third body is outside the circle just defined). Here we must clarify that the tolerances $\delta_1$ and $\delta_2$ were included in the escape criteria in an attempt to avoid that the unstable Lyapunov orbits are incorrectly classified as escaping orbits.

In our numerical integrations the maximum time $t_f$ employed is 5000 dtu (dimensionless time units), corresponding to about 64.315 yr. Usually, the vast majority of orbits need considerable less time to escape from the system (obviously, the numerical integration is effectively ended when an orbit moves outside the system's disk and escapes). Nevertheless, we decided to use such a long integration time just to be sure that all orbits have enough time in order to escape. Remember, that there are the so called ``sticky orbits" which behave as regular ones during long periods of time. Here we should clarify, that orbits which do not escape or collide to Titan after a numerical integration of $5000$ dtu are considered as bounded regular orbits.

The equations of motion (\ref{eqmot}) for the initial conditions of all orbits are forwarded integrated using a double precision Bulirsch-Stoer \verb!FORTRAN 77! algorithm (e.g., \citet{PTVF92}) with a small time step of order of $10^{-2}$, which is sufficient enough for the desired accuracy of our computations. Here we should emphasize, that our previous numerical experience suggests that the Bulirsch-Stoer integrator is both faster and more accurate than a double precision Runge-Kutta-Fehlberg algorithm of order 7 with Cash-Karp coefficients. Throughout all our computations, the Jacobi integral (Eq. (\ref{ham})) was conserved better than one part in $10^{-11}$, although for most orbits it was better than one part in $10^{-12}$. For collisional orbits where the test body moves inside a region of radius $10^{-2}$ around the Titan the Lemaitre's global regularization method is applied.

\section{Numerical results}
\label{numres}

The value of the Jacobi constant defines four different cases regarding the structure of the Hill's regions. In particular
\begin{itemize}
  \item Case I: $C > C_1$ All the transport channels around $L_1$ and $L_2$ are closed, so there is only the bounded and the collisional basins.
  \item Case II: $C_1 > C > C_2$ Only the neck around $L_1$ is open. Therefore, the Saturn realm basin joins the two basins of the previous case.
  \item Case III: $C_2 > C > C_3$ Both necks around $L_1$ and $L_2$ are open. Now we have the exterior realm basin along with the three basins of the second case.
  \item Case IV: $C < C_3$ All four basins are present. In this case however, the forbidden regions are reduced $(C_3 > C > C_4)$ or completely absent $(C < C_4)$.
\end{itemize}

\begin{figure*}[!tH]
\centering
\resizebox{0.7\hsize}{!}{\includegraphics{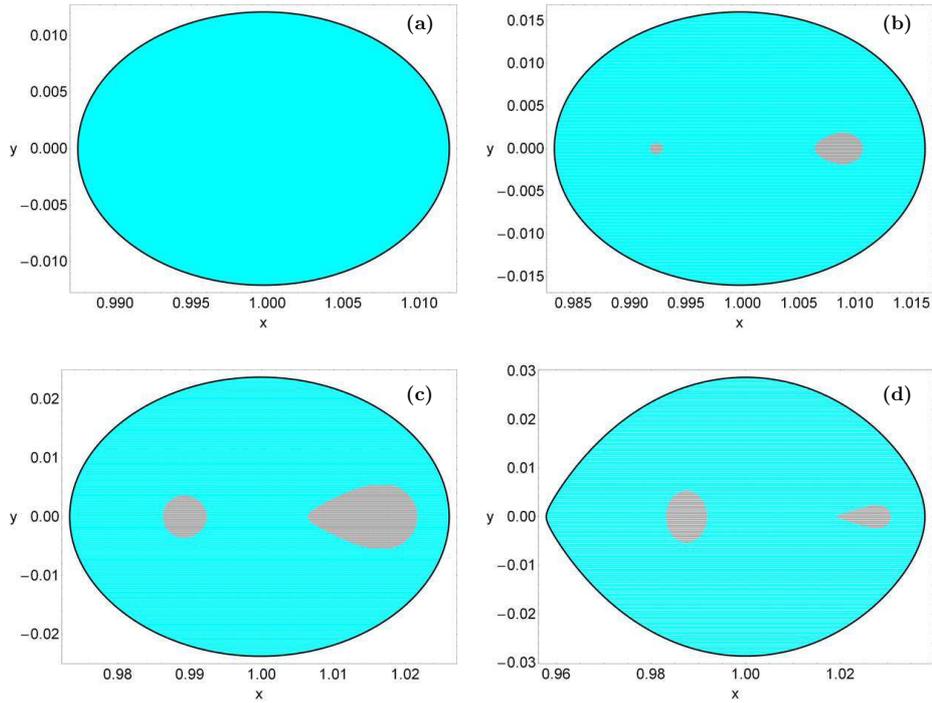}}
\caption{Basin diagrams for Case I. (a-upper left): $C = 3.04$, (b-upper right): $C = 3.03$, (c-lower left): $C = 3.02$ and (d-lower right): $C = 3.01643$. The color code is as follows: bounded basin (gray), collisional basin (cyan).}
\label{hr1}
\end{figure*}

\begin{figure*}[!tH]
\centering
\resizebox{0.7\hsize}{!}{\includegraphics{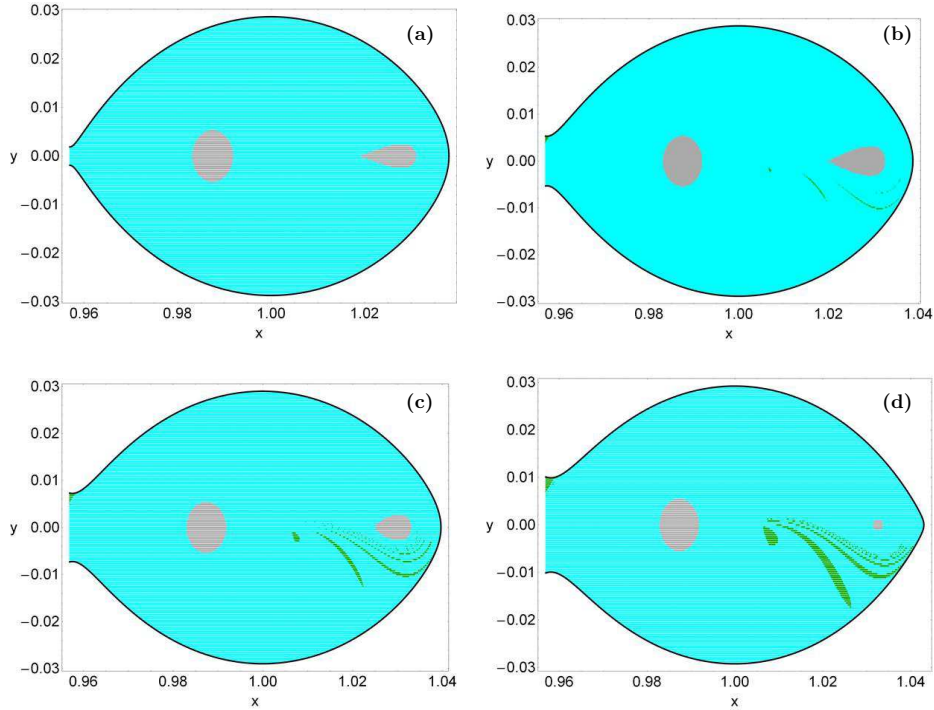}}
\caption{Basin diagrams for Case II. (a-upper left): $C = 3.01641$, (b-upper right): $C = 3.01633$, (c-lower left): $C = 3.01625$ and (d-lower right): $C = 3.01610$. The color code is as follows: bounded basin (gray), collisional basin (cyan), Saturn real basin (green).}
\label{hr2}
\end{figure*}

\begin{figure*}[!tH]
\centering
\resizebox{0.7\hsize}{!}{\includegraphics{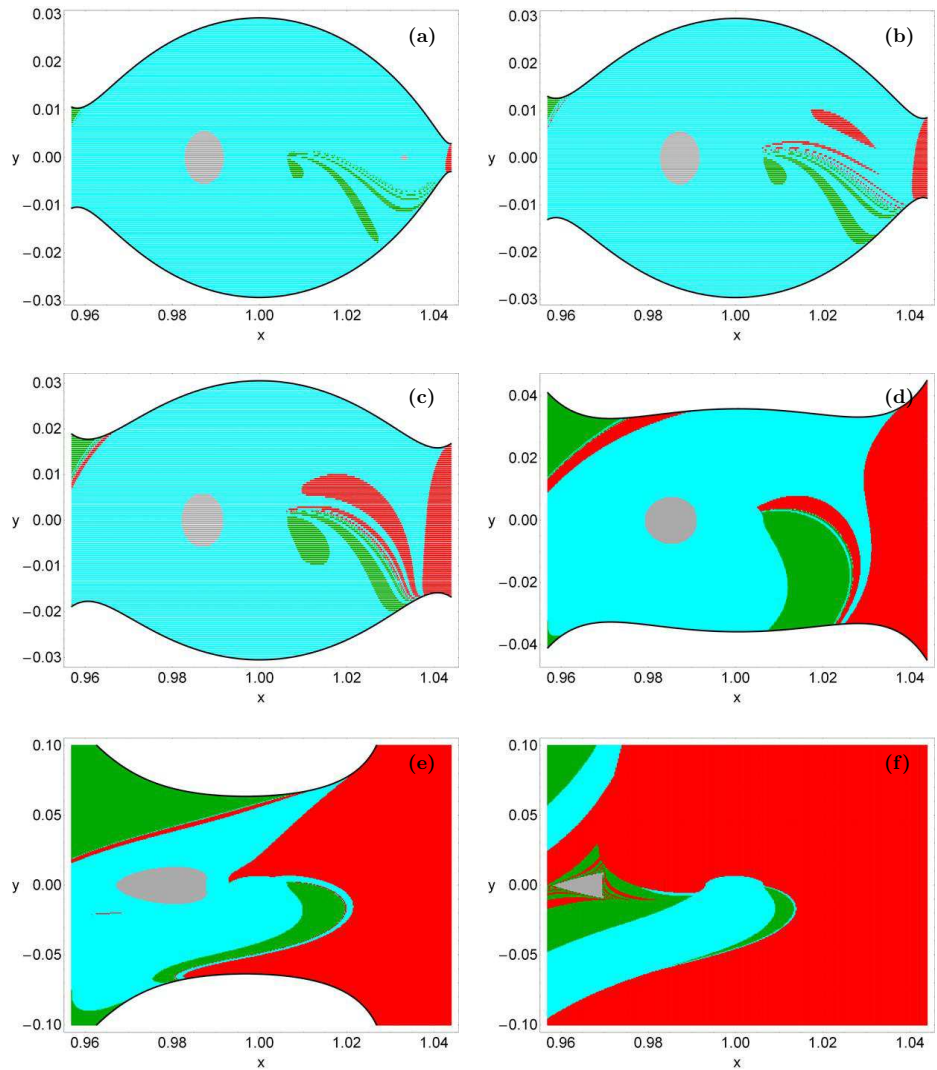}}
\caption{Basin diagrams for Case III. (a): $C = 3.01607$, (b): $C = 3.0159$, (c): $C = 3.0154$, (d): $C = 3.013$, (e): $C = 3.007$, and (f): $C = 3.001$. The color code is as follows: bounded basin (gray), collisional basin (cyan), Saturn realm basin (green), exterior realm basin (red).}
\label{hr3}
\end{figure*}

\begin{figure*}[!tH]
\centering
\resizebox{0.7\hsize}{!}{\includegraphics{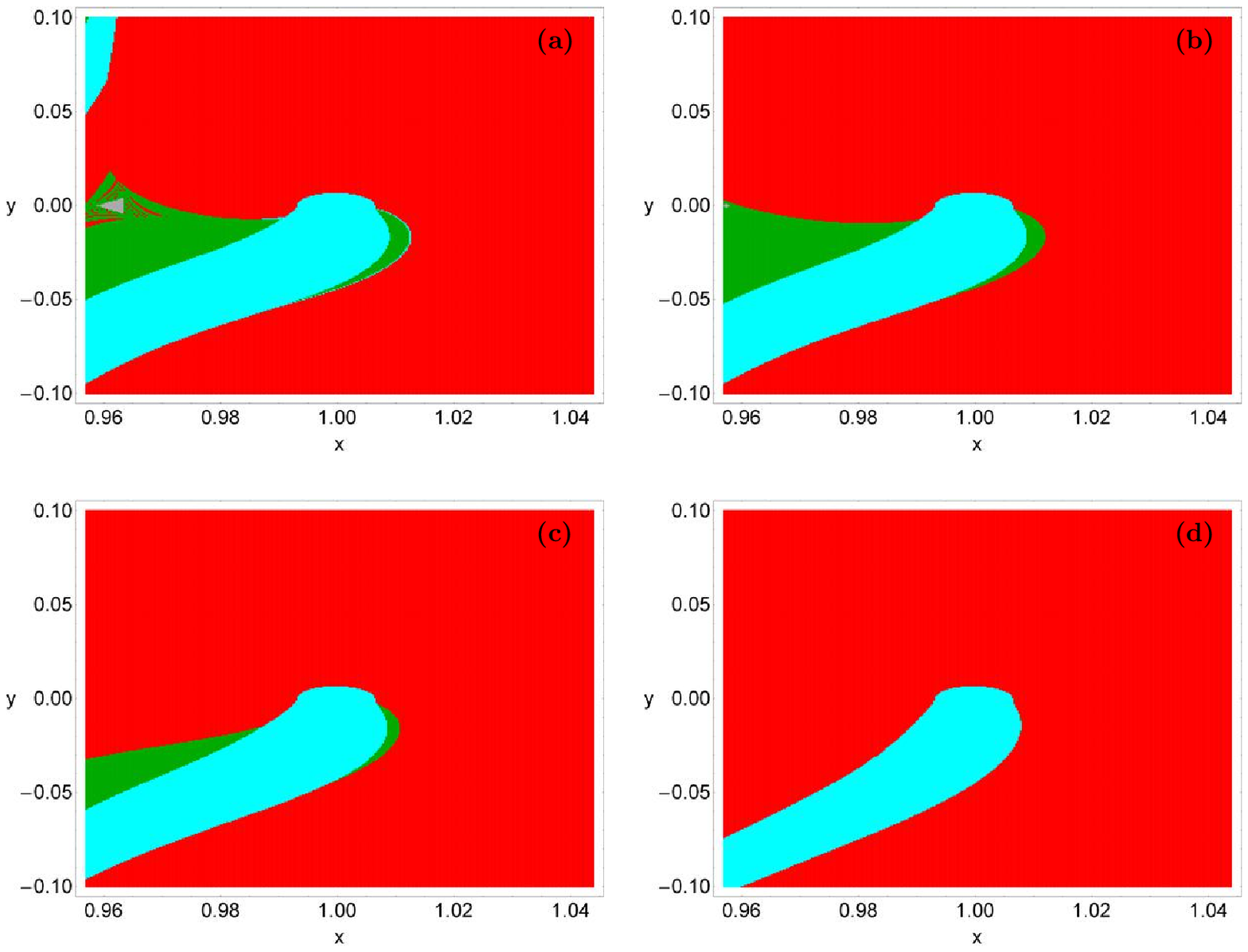}}
\caption{Basin diagrams for Case IV. (a-upper left): $C = 3$, (b-upper right): $C = 2.999$, (c-lower left): $C = 2.995$ and (d-lower right): $C = 2.98$. The color code is the same as in Fig. \ref{hr3}.}
\label{hr4}
\end{figure*}

\subsection{Case I: $C > C_1$}

In Fig. \ref{hr1} we present the basin diagrams for four values of the Jacobi constant. The outermost black solid line is the ZVC which is defined as $\Omega(x,y) = C/2$. Examining these diagrams we see that in Case I there are only bounded and collisional orbits. In particular, for $C = 3.04$, that is an energy level very close to the Titan's surface, all the available configuration space is covered entirely by initial conditions which correspond to collisional orbits\footnote{As in Paper I we also defined initial conditions inside the radius of the Titan which are obviously collisional orbits.}. When $C = 3.03$ two stability islands emerge inside the vast collisional basin. The smaller stability island, located on the left side of the Titan, surrounds a periodic orbit around $P_2$ which is symmetric to a reflection over the $x$-axis $(y = 0)$ and is traveled in counterclockwise sense, hence, prograde with respect to the rotating coordinate system. The larger stability island on the other hand, situated on the right side of the Titan, surrounds a retrograde (clockwise) symmetric periodic orbit around $P_2$. \citet{SS00} proved for the planar Hill problem (see also \citet{dAT14}) that the stability island on the left side of the Titan is much more stable than that on the right side in relation to $C$ variation. As the value of the Jacobi constant decreases we see in Fig. \ref{hr1}c that for $C = 3.02$ both stability islands grow in size. For the lowest value of $C$, for instance $C = 3.01643$ (Fig. \ref{hr1}d), we observe that the size of the stability island located on the left side of the Titan remains almost unperturbed, while the stability region on the right side considerably reduces.

\subsection{Case II: $C_1 > C > C_2$}

When $C < C_1$ the neck around $L_1$ opens thus allowing an orbit to enter the Saturn realm. Indeed for $C = 3.01641$ we see in Fig. \ref{hr2}a that there is an escape channel in the left side of the ZVC. However, even though that the escape is possible we did not detected any initial conditions of orbits that escape through $L_1$ and therefore, the configuration plane is still composed of only bounded and collisional initial conditions. Perhaps for such values of $C$ the width of the escape channel is too small, so all orbits collide to the Titan's surface before they manage to escape. For $C = 3.01633$ the width of the escape channel doubles in size and one can observe the presence of thin filaments of escape basins around the stability island located in the right side of the Titan. As the value of $C$ decreases the width of the escape channel increases thus allowing more and more orbits to escape to the Saturn realm. Apart from the increase to the escape basins we observe that the stability island in the right side of the Titan reduces in size. At the lowest value of the Jacobi constant in Case II, that is $C = 3.01610$, we identify in Fig. \ref{hr2}d well defined escape basins inside the vast collisional basin, while on the right side of the diagram, only a minor bounded basin is present.

Here it should be emphasized that the numerical integration is stopped as soon as the third body passes $L_1$ thus entering the Saturn realm. However, if we do not stop the numerical integration some orbits that initially entered the Saturn realm may return to the Titan region, or crash into on of the primary bodies, or even enter the exterior region and then escape to infinity. In our calculations we follow the approach used in \citet{dAT14} and we consider an orbit to escape to Saturn realm if the test body passes $L_1$ even if its true asymptotic behaviour at very long time limit is different.

\subsection{Case III: $C_2 > C > C_3$}

This case constitutes the Hill's regions with the most interest in point of view of planetary systems. Furthermore, it has many practical applications, such as orbit determination of spacecraft mainly based on many-models and also the phenomenon of temporary capture of comets or asteroids around planets in our solar system. The exterior realm rises when $C < C_2$. In Fig. \ref{hr3}a we see that for $C = 3.01607$ both necks around $L_1$ and $L_2$ are open. Once more, the vast majority of the configuration $(x,y)$ space is covered by initial conditions of orbits which collide to the Titan. It should be noticed that a small escape basin corresponding to the exterior realm is located at the right outer part of the plane very close to $L_2$. When $C = 3.0159$ it is seen in Fig. \ref{hr3}b that additional exterior realm basins emerge inside the collisional basin, while the stability island in the right side of the Titan is no longer present. As we proceed to smaller values of the Jacobi constat (see Figs. \ref{hr3}(c-e)) the extent of the exterior realm basins grows significantly, while the collisional basin shrinks thus confined mainly near the central region of the plane. In particular we see in Fig. \ref{hr3}e that when $C = 3.007$ about 45\% of the available configuration plane is occupied by initial conditions of orbits that escape through $L_2$ to the exterior region. Finally when $C = 3.001$ it is evident that the exterior realm basin take over the configuration plane, leading to the predominance of this type of orbits at the end of Case III. Moreover, the extent of the collisional and the Saturn realm basins is considerably reduced, while a small stability island in the left side of the Titan survives.

\subsection{Case IV: $C < C_3$}

In this last case, there are two subintervals delimited by two critical values of the jacobi constant. In particular, at $C = C_3$ the neck associated with the Lagrangian point $L_3$ opens. Therefore, in the first subcase, $C_3 > C > C_4$ the ZVC and the forbidden regions are still present on the configuration space, while for $C < C_4$ the forbidden regions disappears and all the $(x,y)$ plane is available for motion. Fig. \ref{hr4}a shows the structure of the configuration plane for $C = 3$. We observe that about 80\% of the plane is covered by initial conditions of orbits corresponding to escape through $L_2$. With a much closer look we can identify a small stability island of bounded motion in the left side of the Titan. It should be pointed out that the vicinity of this stability region is highly fractal. This stability island is no longer present in Fig. \ref{hr4}b where $C = 2.999$ thus the configuration plane consists entirely of escape and collisional basins. As we proceed to lower values of the Jacobi constant the extent of the Saturn realm basin is reduced, while that of the collisional basins remains almost unperturbed. At the lowest value of $C$ studied, that is $C = 2.98$, it is seen in Fig. \ref{hr4}d that the Saturn realm basin has disappeared and about 80\% of the $(x,y)$ plane is occupied by initial conditions composing the unified exterior realm basin. We note that the collisional basin has a stream-like shape which is initiated at the center of the plane around the radius of the Titan.

\subsection{An overview analysis}

\begin{figure}[!tH]
\includegraphics[width=\hsize]{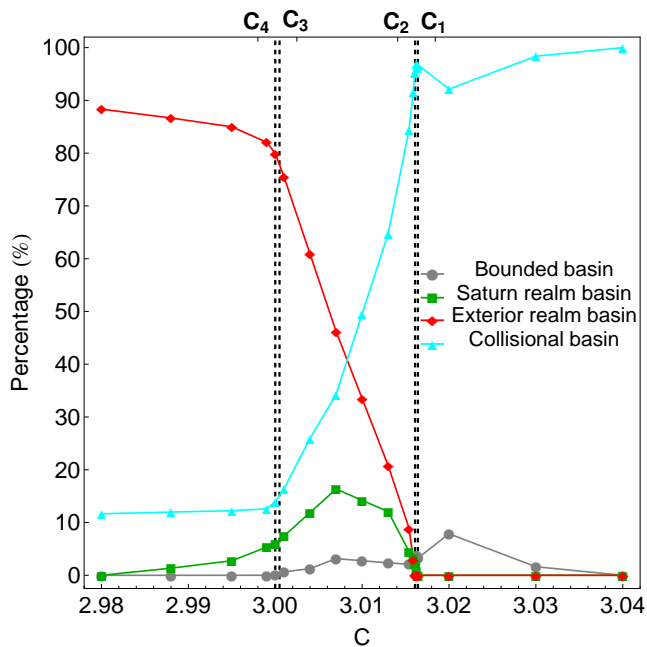}
\caption{Evolution of the percentage of the initial conditions of each considered basin as a function of the Jacobi constant. The vertical dashed black lines indicate the four critical values of $C$.}
\label{percs1}
\end{figure}

It would be very informative to monitor the evolution of the percentages of all types of orbits as a function of the Jacobi constant. Fig. \ref{percs1} shows such a diagram, where the vertical dashed black lines indicate the four critical values of $C$. We observe that at high enough values of the Jacobi constat, that is for $C > 3.02$, the collisional set is the most populated family occupying more than 90\% of the configuration space. However for $C < C_2$ the percentage of collisional orbits exhibits a sharp decrease until $C = C_4$, while for lower values of $C$ it saturates around 12\%. As the neck through $L_1$ opens for $C < C_1$ the rate of escaping orbit to the Saturn realm increases until $C = 3.06$. For lower values of $C$ the trend is reversed and for low enough values of the Jacobi constant $(C < 2.98)$ it completely vanishes. For $C < C_2$ the second channel around $L_2$ also opens and it is seen that the percentage of orbits escaping to the exterior realm displays a rapid increase until $C = C_4$. For lower values of $C$ the slope of the curve changes and the increase is not so dramatic. Nevertheless, when the forbidden regions disappear from the $(x,y)$ plane the exterior realm basin dominates. As for the bounded basin it is evident that it is by far the most weak type of orbits. For $C > C_1$ the rate of bounded orbits has a mean value around 7.5\%, in the interval $C_2 < C < C_4$ its mean value drops to 4.5\%, while for lower values of $C$ this percentage is effectively zero because as we seen in Fig. \ref{hr4}(a-d) the stability island in the left side of the Titan disappears.

\begin{figure*}[!tH]
\centering
\resizebox{0.7\hsize}{!}{\includegraphics{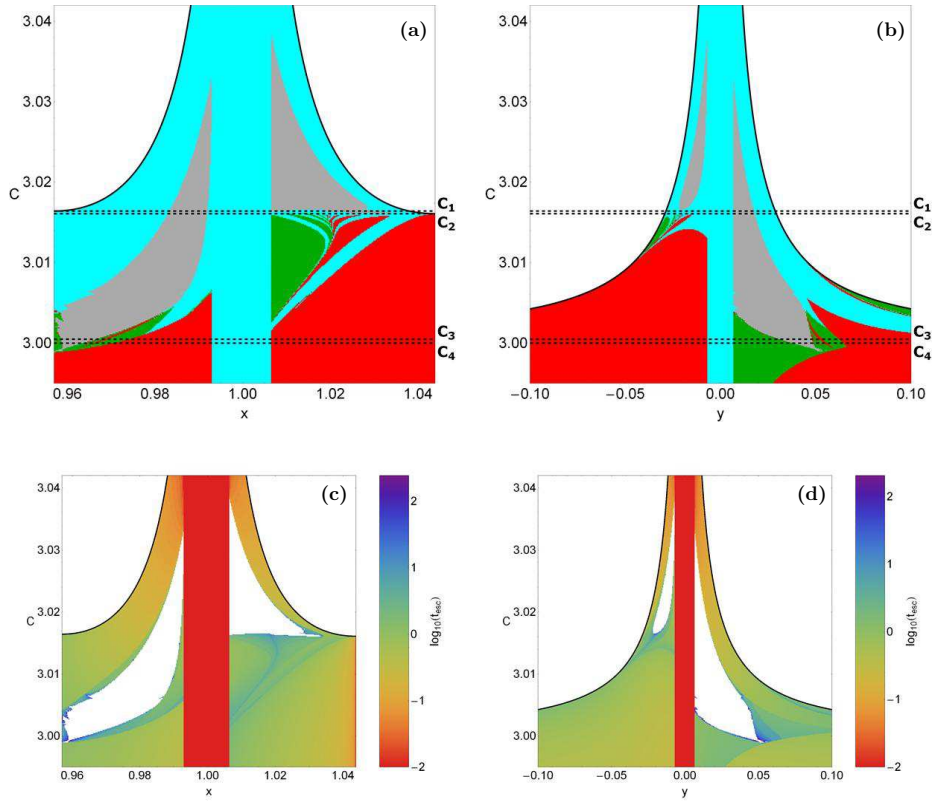}}
\caption{Orbital structure of the (a-upper left): $(x,C)$-plane; and (b-upper right): $(y,C)$-plane; (c-lower left and d-lower right): the distribution of the corresponding escape and collisional times of the orbits. The color code is the same as in Fig. \ref{hr3}.}
\label{xyc}
\end{figure*}

The color-coded planes in the configuration $(x,y)$ space provide sufficient information on the phase space mixing however, for only a fixed value of the Jacobi constant and also for orbits that traverse the surface of section either directly (progradely) or retrogradely. H\'{e}non back in the late 60s \citep{H69}, in order to overcome these limitations introduced a new type of plane which can provide information about areas of bounded and unbounded motion using the section $y = \dot{x} = 0$, $\dot{y} > 0$ (see also \citet{BBS08}). In other words, all the initial conditions of the orbits of the test particles are launched from the $x$-axis with $x = x_0$, parallel to the $y$-axis $(y = 0)$. Consequently, in contrast to the previously discussed type of plane, only orbits with pericenters on the $x$-axis are included and therefore, the value of the Jacobi constant $C$ can now be used as an ordinate. In this way, we can monitor how the variation on $C$ influences the overall orbital structure the Saturn-Titan system using a continuous spectrum of Jacobi constant values rather than few discrete levels. In Fig. \ref{xyc}a we present the orbital structure of the $(x,C)$ plane when $C \in [2.995,3.042]$. The four critical values of $C$ are indicated by the horizontal dashed black lines.

\begin{figure*}[!tH]
\centering
\resizebox{0.9\hsize}{!}{\includegraphics{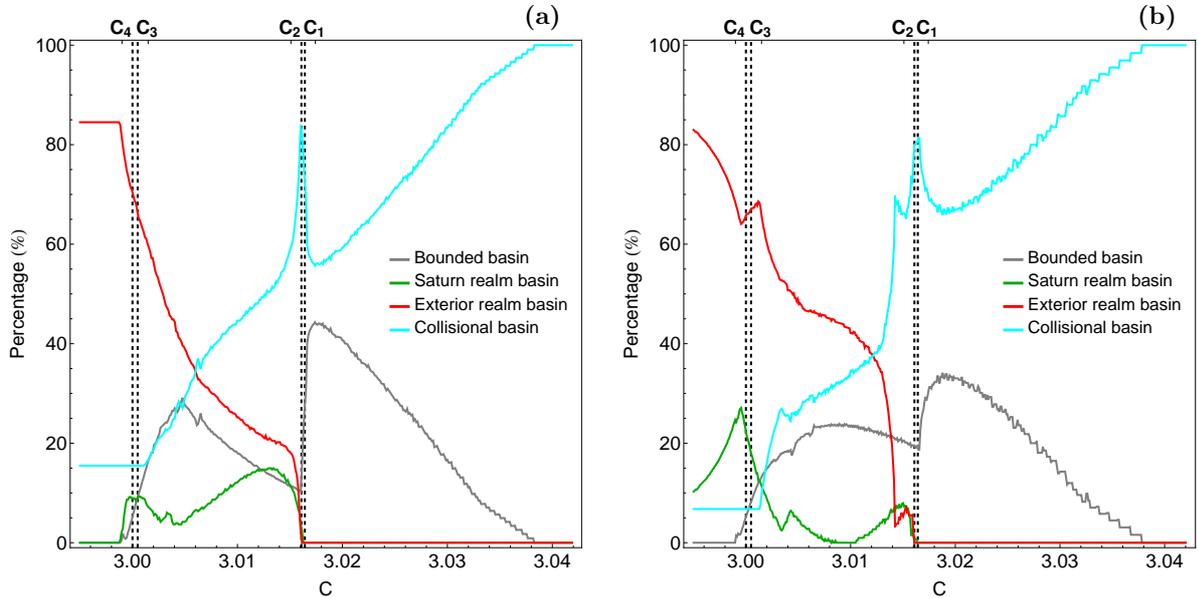}}
\caption{Evolution of the percentages of all types of orbits on the (a-left): $(x,C)$-plane and (b-right): $(y,C)$-plane as a function of the Jacobi constant. The vertical dashed black lines denote the four critical values of $C$.}
\label{percs2}
\end{figure*}

The two stability islands of prograde and retrograde motion in both sides of the Titan are now visible. Here it should be emphasized that around the center of the Titan there is a horizontal lane of collisional orbits. This lane is composed of initial conditions of orbits started inside the radius of the Titan and therefore they are considered to collide with the smaller primary. Below the right stability island there is a fractal mixture of basins corresponding to escaping orbits to the Saturn and the exterior realm. Once more it is evident that for low values of the Jacobi constant the vast majority of the orbits escape to the exterior realm through $L_2$, while for high enough values of $C$ on the other hand all the initial conditions lead to collisional motion. Furthermore, the diagram shown in Fig. \ref{xyc}a show us how the fractality of the several basin boundaries strongly varies not only as a function of the Jacobi constant but also of the spatial variable. In particular, one can observe a very interesting phenomenon. Being more precise we see that the fractality of the basin boundaries, which is related to the unpredictability, migrates from the upper right side of the Titan for relatively high $C$ (or in other words low energy) to the lower left side of the smaller primary for low values of the jacobi constant (high values of the energy).

In Fig. \ref{xyc}c we illustrate how the escape and crash times of orbits are distributed on the $(x,C)$ plane. Light reddish colors correspond to fast escaping/collional orbits, dark blue/purple colors indicate large escape/collional rates, while white color denote stability islands of bounded motion. We note that the scale on the color bar is logarithmic. Inspecting the spatial distribution of various different ranges of the escape time, we are able to associate medium escape time with the stable manifold of a non-attracting chaotic invariant set, which is spread out throughout this region of the chaotic sea, while the largest escape time values on the other hand, are linked with sticky motion around the boundaries of the stability islands. As for the collision time we see that orbits with initial conditions very close to the vicinity of the center of the Titan and for relatively high values of $C$ $(C > 3.03)$ collide with it almost immediately, within the first time step of the numerical integration.

In order to obtain a more complete view on how the nature of orbits in the Saturn-Titan system is influenced by the energy parameter, we follow a similar numerical approach to that explained before but now all orbits are initiated from the vertical $y$-axis with $y = y_0$. In particular, this time we use the section $x = \dot{y} = 0$, $\dot{x} > 0$, launching orbits parallel to the $x$-axis. This allow us to construct again a 2D plane in which the $y$ coordinate of orbits is the abscissa, while the value of the Jacobi constant is the ordinate. The orbital structure of the $(y,C)$-plane for the same range of values of $C$ is shown in Fig. \ref{xyc}b. The black solid line is the limiting curve which distinguishes between regions of allowed and forbidden motion and is defined as
\begin{equation}
f_L(y,C) = \Omega(x = 0,y) = C/2.
\label{zvc}
\end{equation}
A very complicated orbital structure is reveled in the $(y,C)$-plane which however, in general terms, is very similar with that of the $(x,C)$-plane. One may observe that for $C < C_4$ the forbidden regions of motion around $L_4$ and $L_5$ completely disappear. The corresponding escape and collisional times of the orbits are given in Fig. \ref{xyc}d. Looking carefully Figs. \ref{xyc}(c-d) we may conclude that the smallest escape/collisonal periods correspond to orbits with initial conditions inside the escape/collison basins, while orbits initiated in the fractal regions of the planes or near the boundaries of stability islands have the highest escape rates.

The evolution of the percentages of all types of orbits on the $(x,C)$ and $(y,C)$ planes as a function of the Jacobi constant are presented in Fig. \ref{percs2}(a-b). We observe that the patterns of both diagrams are very similar. At this points it should be very informative to briefly discuss the evolution of the percentages. We will analyze the patterns of Fig. \ref{percs2}a; those shown in Fig. \ref{percs2}b are quite similar. When $C > 3.04$ all the tested initial conditions correspond to collisional orbits. As the value of $C$ decreases however, the rate of collisional orbits decreases almost linearly until $C_1$. In the interval $C_2 < C < C_1$ we observe a sudden peak at about 80\%, while for the interval $C_3 < C < C_2$ the linear decrease continues until $C_3$. For all lower values of the Jacobi integral the percentage of collisional orbits remains constant and about 15\%. For $C < C_1$ the neck near $L_1$ opens and the orbits may enter the Saturn realm. Indeed, we see that in the interval $C_3 < C < C_1$ the rate of escaping orbits trough $L_1$ exhibits fluctuations between 5\% and 15\%, while for $C < C_4$ it suddenly vanishes. The channel leading to escape to the exterior realm opens for $C < C_2$, where the corresponding percentage grows rapidly until about $C_4$. For lower values of $C$ it stabilizes around 84\%. Finally the rate of the bounded basin increases from 0\% at $C = 3.038$ to about 45\% at $C_1$. Inside the interval $C_2 < C < C_1$ there is a sudden reduction, while the growth continues up to $C = 3.005$ $(30\%)$. For smaller values of $C$, or in other words for higher energy levels, the percentage of bounded orbits reduces and for $C < C_4$ it tends asymptotically to zero. Therefore, one may reasonably conclude that at low energy levels which correspond to close ZVCs around the Titan collisional motion dominates, while at high energy levels on the other hand the vast majority of orbits escape to the exterior region.

\section{Discussion and conclusions}
\label{disc}

The planar version of the circular restricted three-body problem was used in this contribution in order to numerically investigate the orbital dynamics of a small body (spacecraft, comet or asteroid) under the influence of the potential of the Saturn-Titan system. All the initial conditions of orbits were initiated in the vicinity of the Titan which was our scattering region. We managed to determine four basins corresponding to: (i) bounded orbits around the Titan, (ii) escaping orbits to the Saturn realm through $L_1$, (iii) escaping orbit to the exterior realm through $L_2$, (iv) leaking orbits due to collisions with the surface of the Titan. The orbital structure of the basins was monitored by varying the value of the Jacobi constant. As far as we know, this is the first time that the orbital content in the vicinity of the Titan is explored in such a detailed and systematic way.

We believe that our results could be useful in space mission design. Nevertheless, we have to stress out that in real space missions there are several types of plane perturbations mainly due to the presence of other heavy celestial bodies (e.g., planets). Near the vicinity of Titan (which was the scattering region in our work) however, all these perturbations are extremely weak and therefore negligible, so we may claim that the presented numerical results are structurally stable against out of plane perturbations.

For several values of the Jacobi integral we defined dense, uniform grids of $1024 \times 1024$ initial conditions $(x_0,y_0)$ regularly distributed in the area allowed by the Jacobi constant on the configuration $(x,y)$ space and then we forwarded integrated them. For the numerical integration of the orbits in every grid, we needed about between 1 minute and 5 hours of CPU time on a Pentium Dual-Core 2.2 GHz PC, depending on the escape, collisional and bounded rates of orbits in each case. For each initial condition, the maximum time of the numerical integration was set to be equal to 5000 dtu however, when a particle escaped or collided the numerical integration was effectively ended and proceeded to the next available initial condition.

Our work contains quantitative information regarding the orbital dynamics in the Saturn-Titan. The main numerical results can be summarized as follows:
\begin{enumerate}
 \item In all examined cases, regions of initial conditions leading to escape in a given direction or to collision with the smaller primary were found to exist. The several escape/collision basins are very intricately interwoven and they appear either as thin elongated spiral bands or as well-defined broad regions.
 \item It was found that for high values of the Jacobi constant, or in other words for low energy levels, the configuration space is dominated by collisional orbits, while some stability islands are also present.
 \item We observed that for low values of the Jacobi constant which correspond to high values of the total orbital energy the vast majority of the orbits escape to the exterior realm, while the rest of the $(x,y)$ plane is occupied by collisional orbits.
 \item It was detected that the average collisional time increases with increasing $C$, while on the other hand, the average escape time, in general terms, diminishes with decreasing $C$.
 \item Our calculations revealed that inside the basins the shortest escape and collision rates of the orbits had been measured. On the contrary, the longest escape/collision rates correspond to initial conditions located near the boundaries between the basins or in the vicinity of the stability islands.
\end{enumerate}

Taking into account the detailed and systematic analysis we may conclude that out numerical task has been successfully completed. We hope that the present results to be useful in the active field of orbital dynamics in the restricted three-body problem. Since our outcomes are encouraging, it is in our future plans to expand our investigation into three dimensions and also in other planetary systems.

\section*{Acknowledgments}

I would like to thank the anonymous referee for the careful reading of the manuscript and for all the apt suggestions and comments which allowed us to improve both the quality and the clarity of the paper.

\end{document}